\documentclass[preprint,nofootinbib,amsmath,amssymb]{revtex4}

\usepackage[dvips]{graphicx}

\usepackage{amssymb} 

\begin{document}

\title{Solar-System Constraints on $f(R)$ Chameleon Gravity}

\author{Je-An~Gu$^{1}$}
\email{jagu@ntu.edu.tw} %
\author{Wei-Ting Lin$^{1,2}$}
\email{r97222021@ntu.edu.tw}
\affiliation{
${}^1$Leung Center for Cosmology and Particle Astrophysics, National Taiwan University, Taipei 10617, Taiwan, R.O.C.\\
${}^2$Department of Physics, National Taiwan University, Taipei 10617, Taiwan, R.O.C.
}

\date{\today}

\begin{abstract} 
We investigate the solar-system constraint on the $f(R)$ theory
of modified gravity with chameleon mechanism, where $f(R)$
represents the deviation from general relativity in the gravity
action. We obtain a stringent bound to a general, non-constant
deviation function $f(R)$: $-10^{-15} \lesssim df/dR < 0$ when
$R/H_0^2 \sim 3 \times 10^5$, and a loose bound: 
$0 < R \, d^2f/dR^2 < 2/5$ when $R/H_0^2 \gtrsim 3 \times 10^5$, 
by requiring the thin-shell condition in the solar system, 
particularly in the atmosphere of the Earth. 
These bounds can be conveniently utilized to test the $f(R)$ models 
with given functional forms of $f(R)$ and to obtain the constraints on 
the parameters therein. For demonstration we apply these bounds to 
several widely considered $f(R)$ models.
\end{abstract}


\maketitle

\section{Introduction} \label{sec:introduction}

The discovery of the cosmic acceleration at the present epoch
\cite{Perlmutter:1998np,Riess:1998cb} indicates the existence
of repulsive gravity that dominates the present universe at large scales. 
It may be an indication of an energy source of anti-gravity 
or a sign of the modification of the gravity theory.
Along the latter direction the $f(R)$ theory of modified gravity
\cite{Capozziello:2003tk,Carroll:2003wy,Nojiri:2003ft,Appleby:2007vb,Hu:2007nk,Starobinsky:2007hu,Nojiri:2007as,Tsujikawa:2007xu,Nojiri:2007cq,Cognola:2007zu,Linder:2009jz}
(for a review, see \cite{DeFelice:2010aj}) 
has been proposed as a possible explanation of the cosmic acceleration, 
where the gravity is described by a function of the Ricci scalar $R$ in the action. 
On the other hand, the $f(R)$ theory can be treated as a simple way of
modeling the possible deviation from Einstein's general relativity (GR).

As an essence of cosmology, $f(R)$ gravity needs to
pass the cosmological test
\cite{Amendola:2007nt,Tsujikawa:2008uc,Serra:2009kp,Daniel:2009kr,Guzik:2009cm,Tsujikawa:2009ku,Giannantonio:2009gi,Daniel:2010ky,Bean:2010zq,LinGuChen:2010}
involving the observations of the cosmic expansion and the
cosmic structure formation. As a gravity theory, it needs to
pass the local gravity test that in general gives the
most stringent constraint on modified gravity so far
\cite{Chiba:2006jp,Faulkner:2006ub,Amendola:2007nt,Capozziello:2007eu,Tsujikawa:2008uc,Bisabr:2009ee}. 
It has been pointed out that $f(R)$ gravity, as a special
case of the scalar-tensor theory, may pass the solar-system
test with the help of the ``chameleon mechanism''
\cite{Khoury:2003aq,Khoury:2003rn}, meanwhile driving the
late-time cosmic acceleration. In the chameleon mechanism the
scalar field of the scalar-tensor theory can behave differently
in different environments, depending on the ambient mass
density. This feature makes it possible to have significant
deviations from GR at the cosmological scales at late times and
meanwhile have tiny deviations both in the solar system at
present and at all scales at early times.

In this paper we investigate the solar-system constraint on the
$f(R)$ gravity with the chameleon mechanism.\footnote{We 
consider the metric formalism of $f(R)$ gravity.}
As a result, we obtain the following constraints on a
general, non-constant function $f(R)$ that represents the
deviation from GR in the gravity action.
\begin{eqnarray}
-10^{-15} \lesssim df/dR < 0
&\textrm{ when }& R/H_0^2 \sim 3 \times 10^5 , \\
0 < R \, d^2f/dR^2 < 2/5
&\textrm{ when }& R/H_0^2 \gtrsim 3 \times 10^5 .
\end{eqnarray}
These constraints will be derived in Sec.\ \ref{sec:constraints}. 
Before that, in Sec.\ \ref{sec:f(R)-chameleon} we will introduce
$f(R)$ gravity, the chameleon mechanism and the thin-shell condition, 
and in Sec.\ \ref{sec:thin-shell condition} elaborate on 
the thin-shell parameter in $f(R)$ gravity 
and obtain its relation to $f(R)$, with which 
the solar-system constraint on the thin-shell parameter
can be transferred to that on $f(R)$.

\section{$f(R)$ Gravity with Chameleon Mechanism} \label{sec:f(R)-chameleon}


We consider the $f(R)$ theory of modified gravity with the
action,\footnote{We use the convention, $\{-,+,+,+\}$, for the
metric signature.}
\begin{equation}
S = \frac{1}{2\kappa^{2}}\int d^4 x\sqrt{-g}
\left[R+f(R)\right] + S_{m} \left( g_{\mu \nu} , \Psi_{m} \right) ,
\label{eq:f(R)action}
\end{equation}
where $f(R)$ is a function of the Ricci scalar $R$ and
represents the deviation from GR,  $\kappa$ is the gravitational 
constant, $S_{m}$ the matter action and $\Psi_m$ the matter field. 
This theory can be transformed to a scalar-tensor theory 
in the Einstein frame via a conformal transformation,
\begin{equation}
\tilde{g}_{\mu \nu} = \left( 1 + f_R \right) g_{\mu \nu}
= e^{-2\beta\kappa\phi}g_{\mu\nu} \, , \quad
\beta = -1/\sqrt{6} \, ,
\label{eq:conformal transf}
\end{equation}
where $f_{R} \equiv df/dR$. 
The resultant action in the Einstein frame is
\begin{equation}
S_E = \int d^4 x \sqrt{-\tilde{g}} \left[
\frac{1}{2\kappa^{2}} \tilde{R}
- \frac{1}{2} \tilde{g}^{\mu \nu} \left( \partial_{\mu} \phi \right)
\left( \partial_{\nu} \phi \right) - V(\phi) \right]
+ S_{m} \left( e^{2\beta \kappa \phi} \tilde{g}_{\mu \nu} ,
\Psi_{m} \right) , \label{eq:action in Einstein frame}
\end{equation}
where
\begin{eqnarray}
\phi &=& -\frac{1}{2\beta\kappa} \ln \left[ 1+ f_{R}(R) \right] \, , \label{eq:phi}\\
V(\phi) &=& \frac{Rf_{R}(R)-f(R)}{2\kappa^2 \left[1+f_{R}(R)\right]^2} \, , \label{eq:V(phi)} \\
\tilde{R} &=& R \left[ \tilde{g}_{\mu \nu} \right] .
\end{eqnarray}
The notation about the derivatives of $f(R)$,  
$f_{R} \equiv df/dR$ and $f_{RR} \equiv d^2 f/dR^2$, 
will be used in the remainder of this paper.

The potential $V$ in Eq.\ (\ref{eq:V(phi)}) is a function
of the scalar field $\phi$, provided that the
Ricci scalar $R$ is a function of $\phi$ given by Eq.\
(\ref{eq:phi}) with a well-defined inverse function $f_R^{-1}$. 
In the action $S_E$ the scalar field acquires an
additional coupling to matter through the Jordan-frame metric
tensor $g_{\mu \nu}$ in the matter action $S_m$. 
This additional coupling is the key to the chameleon mechanism.
It makes the scalar field $\phi$ behave differently in different 
environments, depending on the ambient mass density.


To study the $f(R)$ chameleon gravity with the action $S_E$, 
for simplicity we consider a spherically symmetric system,
especially a sphere with different constant mass densities
inside and outside the sphere. In the following we will present
the field equation and the solution for the scalar field
$\phi(r)$, where $r$ is the physical distance from the center
of the system.

For a spherically symmetric space-time with 
a mass density distribution 
$\rho^{\ast}(r)$, the field equation is
\begin{equation}
\frac{d^2\phi}{dr^2} + \frac{2}{r}\frac{d\phi}{dr}
= \frac{\partial  V_\textrm{eff}(\phi,\rho^{\ast})}{d\phi} ,
\label{eq:phi field eqn}
\end{equation}
where the effective potential
\begin{equation}
V_\textrm{eff} (\phi,\rho^{\ast}) \equiv
V(\phi) + e^{\beta \kappa \phi} \rho^{\ast} .
\end{equation}
The mass density $\rho^{\ast}$ is a conserved quantity in the
Einstein frame. By $\rho^{\ast} = e^{3\beta \kappa \phi} \rho$,
it is related to the mass density $\rho$ 
in the $f(R)$ gravity with the action $S$. 

To satisfy the constraints from the solar-system experiments,
the effective potential $V_\textrm{eff}$ should have a minimum
and should be steep around the minimum when the mass density
$\rho^{\ast}$ is equal to those 
in the solar system, including the regions inside the Earth
($\rho^{\ast}_{\oplus} \simeq 5.5 \textrm{ g}/\textrm{cm}^3$),
in the atmosphere ($\rho^{\ast}_\textrm{atm} \sim 10^{-3}
\textrm{ g}/\textrm{cm}^3$) and outside the atmosphere
($\rho^{\ast}_{G} \sim 10^{-24} \textrm{ g}/\textrm{cm}^3$).
Let $\phi_m$ denote the location of the potential minimum and
$m$ the mass of $\phi$, i.e.,
\begin{equation}
\partial_{\phi} V_\textrm{eff} \left( \phi_m , \rho^{\ast} \right) = 0 ,
\end{equation}
\begin{equation}
m^2 \left( \rho^{\ast} \right)
\equiv \partial_{\phi}^2 V_\textrm{eff} (\phi_m , \rho^{\ast}) .
\end{equation}
The basic condition raised above is then recast by the
requirements: (1) the existence of $\phi_m$, (2) positive $m^2$, 
and (3) large $m$ (i.e.\ short Compton wavelength), when
$\rho^{\ast} \sim \rho^{\ast}_{\oplus}$,
$\rho^{\ast}_\textrm{atm}$, $\rho^{\ast}_{G}$, etc.

We consider a sphere with the radius $r_s$ and with different
constant mass densities, $\rho^{\ast}_\textrm{in}$ and
$\rho^{\ast}_\textrm{out}$, inside and outside the sphere. 
The locations of the potential minima $\phi_c$
and $\phi_{\infty}$, the masses $m_\textrm{in}$ and
$m_\textrm{out}$, the total mass of the sphere $M_s$ and the
Newtonian potential $\Phi_s$ at $r_s$ (i.e.\ on the surface of
the sphere) are defined as follows.
\begin{equation}
\partial_{\phi} V_\textrm{eff} \left( \phi_c , \rho^{\ast}_\textrm{in} \right) = 0 , %
\quad %
\partial_{\phi} V_\textrm{eff} \left( \phi_{\infty} , \rho^{\ast}_\textrm{out} \right) = 0 ,
\label{eq:minima location in and out}
\end{equation}
\begin{equation}
m_\textrm{in}^2 \equiv \partial_{\phi}^2 V_\textrm{eff} \left( \phi_c,\rho^{\ast}_\textrm{in} \right) , %
\quad %
m_\textrm{out}^2 \equiv \partial_{\phi}^2 V_\textrm{eff} \left( \phi_{\infty},\rho^{\ast}_\textrm{out} \right) ,
\label{eq:masses in and out}
\end{equation}
\begin{equation}
M_s \equiv \frac{4\pi}{3} r_s^3 \rho^{\ast}_\textrm{in} \, , \quad %
\Phi_s \equiv \frac{\kappa^2}{8\pi} \frac{M_s}{r_s} .
\label{eq:sphere mass and potential}
\end{equation}
The $\phi$ field profile, as a solution of Eq.\ (\ref{eq:phi
field eqn}), is
\begin{equation}
\left\{
\begin{array}{rcl} \label{eq:phi-profile}
\phi \left( r < r_s \right) &\simeq& \phi_c \, , \\ 
\phi \left( r > r_s \right) &\simeq& \displaystyle
    - \left( \frac{\beta\kappa}{4\pi} \right)
      \left( \frac{3\Delta r_s}{r_s} \right)
      \frac{M_s e^{-m_\textrm{out}(r-r_s)}}{r}
    + \phi_{\infty} \, , 
\end{array}
\right.
\end{equation}
when the following thin-shell condition is satisfied.
\begin{equation}
0 < \frac{\Delta r_s}{r_s} \equiv
\frac{\kappa(\phi_{\infty}-\phi_c)}{6\beta\Phi_s} \ll 1 .
\label{eq:thin-shell condition}
\end{equation}

The thin-shell parameter $\Delta r_s / r_s$ is proportional to
the ratio of two potential differences, 
$\kappa \phi_{\infty} - \kappa \phi_c$ and 
$\Phi_s - \Phi_\infty$ [where $\Phi_\infty \equiv \Phi
(r=\infty) \equiv 0$], which respectively relate to the
strength of the fifth force induced by $\phi$ and that of the
Newtonian gravitational force. Thus, roughly speaking, the
thin-shell condition requires the weakness of the fifth force
compared to the Newtonian gravity.

The $\phi$ profile in Eq.\ (\ref{eq:phi-profile})
depends on the thin-shell parameter $\Delta r_s / r_s$ 
that is determined when the function $f(R)$ and the mass densities,
$\rho^{\ast}_\textrm{in}$ and $\rho^{\ast}_\textrm{out}$, are given.
Accordingly, the solar-system bounds to the fifth force can
constrain $\Delta r_s / r_s \,$ and thereby constrain $f(R)$.

\section{Thin-shell parameter in $f(R)$ gravity} \label{sec:thin-shell condition}

Here we will derive the relation between the thin-shell parameter
$\Delta r_s / r_s$ and the function $f(R)$. 
With this relation at hand we can obtain the constraint on 
general $f(R)$ from the solar-system tests of gravity
which give bounds to the thin-shell parameter. 
In addition, 
the requirement of the existence of the potential minima, 
$m_\textrm{in}^2 > 0$ and $m_\textrm{out}^2 > 0$, 
also gives a basic condition of $f(R)$, for which we will
derive the relation between $f(R)$ and the masses, 
$m_\textrm{in}$ and $m_\textrm{out}$.

The thin-shell parameter is related to $f(R)$ through the
locations of the potential minima, $\phi_c$ and
$\phi_{\infty}$, which satisfy Eq.\ (\ref{eq:minima location in
and out}). With $dV/d\phi = (dV/dR) (dR/d\phi)$ and 
$e^{\beta \kappa \phi} = 1/\sqrt{1+f_R}$ we have
\begin{eqnarray}
\partial_{\phi} V_\textrm{eff} \left( \phi , \rho^{\ast} \right)
&=& \frac{dV}{d\phi} + \beta \kappa e^{\beta \kappa \phi} \rho^{\ast} \\
&=& \frac{R+2f-Rf_R}{\sqrt{6}\kappa \left( 1+f_R \right)^2}
+ \frac{\beta \kappa \rho^{\ast}}{\sqrt{1+f_R}} \, ,
\end{eqnarray}
\begin{eqnarray}
\partial_{\phi}^2 V_\textrm{eff} \left( \phi , \rho^{\ast} \right)
&=& \frac{d^2V}{d\phi^2} + \beta^2 \kappa^2 e^{\beta \kappa \phi} \rho^{\ast} \\
&=& \frac{1}{3f_{RR}} - \frac{3R+4f-Rf_R}{3\left( 1+f_R \right)^2}
+ \frac{\beta^2 \kappa^2 \rho^{\ast}}{\sqrt{1+f_R}} \, .
\end{eqnarray}
In the solar system the deviation from GR must be small, i.e.,
$|f| \ll R$ and $|f_R| \ll 1$, which we will use for the approximation
involved in the following derivation. Therefore,
\begin{equation}
\partial_{\phi} V_\textrm{eff}
\simeq \frac{1}{\sqrt{6}\kappa}
\left[
\left( R - \kappa^2 \rho^{\ast} \right)
+ \left( 2f - 3Rf_R + \frac{1}{2} f_R \kappa^2 \rho^{\ast} \right)
\right] ,
\end{equation}
\begin{equation}
\partial_{\phi}^2 V_\textrm{eff} \simeq
\left( \frac{1}{3f_{RR}} - R + \frac{1}{6}\kappa^2 \rho^{\ast} \right)
+ \left( - \frac{4}{3}f + \frac{7}{3}Rf_R - \frac{1}{12} f_R \kappa^2 \rho^{\ast} \right) .
\end{equation}

The location of the potential minimum, $\phi_m$, is given by
$\partial_{\phi} V_\textrm{eff} (\phi_m,\rho^{\ast})$=0, and
therefore satisfies
\begin{equation}
\left[ R + 2f - 5Rf_R/2 \right]_{\phi_m} \simeq \kappa^2 \rho^{\ast} ,
\end{equation}
or, to the lowest order,
\begin{equation}
R \left( \phi_m \right) \simeq \kappa^2 \rho^{\ast} .
\end{equation}
Thus,
\begin{eqnarray}
\phi_m &\simeq&
- \frac{1}{2\beta \kappa} \ln \left[ 1 + f_R (R \simeq \kappa^2
\rho^{\ast}) \right] \\
&\simeq& - \frac{1}{2\beta \kappa} f_R (R \simeq \kappa^2
\rho^{\ast}) .
\end{eqnarray}
We then obtain the formula for the mass,
\begin{equation}
m^2 \left( \rho^{\ast} \right)
= \partial_{\phi}^2 V_\textrm{eff} \left( \phi_m,\rho^{\ast} \right)
\simeq \left[ \left( \frac{1}{3f_{RR}} - \frac{5}{6}R \right)
    + \left( -f + \frac{11}{6}Rf_R \right) \right]_{R\simeq \kappa^2\rho^{\ast}} ,
\label{eq:mass-square-approx}
\end{equation}
and the relation between the thin-shell parameter and $f(R)$,
\begin{equation}
\frac{\Delta r_s}{r_s} \simeq
- \frac{f_R(\kappa^2 \rho^{\ast}_\textrm{out})-f_R(\kappa^2 \rho^{\ast}_\textrm{in})}
{2\Phi_s} \, .
\label{eq:thin-shell parameter-fR}
\end{equation}

In many viable $f(R)$ models, $|f_R(R_1)| \gg |f_R(R_2)|$ when
$R_1 \ll R_2$, in order to fit the cosmic microwave background
(CMB) and the big-bang nucleosynthesis (BBN) observational
results that require the deviation from GR be tiny at early times,
meanwhile generating the cosmic acceleration with a significant
deviation from GR at late times. In this case,
\begin{equation}
\Delta r_s / r_s \simeq - f_R(\kappa^2
\rho^{\ast}_\textrm{out}) / 2\Phi_s
\quad \textrm{when} \quad
\rho^{\ast}_\textrm{out} \ll \rho^{\ast}_\textrm{in} \, .
\label{eq:thin-shell parameter-fR-simple}
\end{equation}
The condition $\rho^{\ast}_\textrm{out} \ll
\rho^{\ast}_\textrm{in}$ is satisfied in many solar-system
experiments.

\section{Solar-System Constraints on $f(R)$ Gravity} \label{sec:constraints}

To fit the solar-system constraints,
the thin-shell condition in Eq.\ (\ref{eq:thin-shell
condition}) needs to be satisfied in the solar system. 
This condition requires (1)
the existence of the minimum of the effective potential
$V_\textrm{eff} (\phi , \rho^{\ast})$ and (2) the smallness of
$|f_R|$.

From Eq.\ (\ref{eq:mass-square-approx}), the existence of the
minimum of $V_\textrm{eff}$, i.e.\ $m^2>0$, entails (to the
lowest order)
\begin{equation}
0 < R f_{RR} < 2/5 \quad \textrm{when}
\quad R \simeq \kappa^2 \rho^{\ast} \gtrsim 3 \times 10^5 H_0^2 ,
\label{eq:fRR constraint}
\end{equation}
where we have considered various environments in the solar
system, with the mass density $\rho^{\ast}$ ranging from
$10^{-24} \textrm{ g}/\textrm{cm}^3$ (Space) to $5.5 \textrm{
g}/\textrm{cm}^3$ (Earth) and accordingly with $R/H_0^2$
ranging from $3 \times 10^5$ to $\mathcal{O}(10^{30})$. 
For simplicity we use $R \gtrsim 3 \times 10^5 H_0^2$ 
in the above expression.

Regarding the thin-shell condition, an upper bound $(\Delta r_s
/ r_s)_\textrm{max}$ of the thin-shell parameter constrains
$f_R$ via Eqs.\ (\ref{eq:thin-shell condition}) and
(\ref{eq:thin-shell parameter-fR}) as follows.
\begin{equation}
0 < f_R(\kappa^2 \rho^{\ast}_\textrm{in}) - f_R(\kappa^2 \rho^{\ast}_\textrm{out})
< 2 \Phi_s \cdot \left( \Delta r_s / r_s \right)_\textrm{max} \ll \Phi_s \, ,
\end{equation}
which gives an upper bound to the $f_R$ variation. In the cases where 
\begin{equation} \label{eq:fR-in-out-condition}
\left| f_R(\kappa^2 \rho^{\ast}_\textrm{in}) \right| \ll
\left| f_R(\kappa^2 \rho^{\ast}_\textrm{out}) \right| \mbox{ for }
\rho^{\ast}_\textrm{in} \gg \rho^{\ast}_\textrm{out} \, , 
\end{equation}
it gives an upper bound to $|f_R(\kappa^2
\rho^{\ast}_\textrm{out})|$:
\begin{equation}
0 < - f_R < 2 \Phi_s \cdot \left( \Delta r_s / r_s \right)_\textrm{max} \ll \Phi_s
\quad \textrm{when} \quad R \simeq \kappa^2 \rho^{\ast}_\textrm{out} \, .
\end{equation}
In the following we will consider the upper bound $(\Delta r_s
/ r_s)_\textrm{max}$ obtained from the solar-system experiments
involving the Sun and the Earth, respectively.

The experimental tests of the post Newtonian parameters 
in the solar system \cite{Will:2005va} give \cite{Capozziello:2007eu}
\begin{equation}
\left( \Delta r_s / r_s \right)_\textrm{max} = 1.15 \times 10^{-5}.
\end{equation}
With $\Phi_{\odot} \simeq 2.12 \times 10^{-6}$ for the Sun and
$\rho^{\ast}_\textrm{out} \simeq \rho^{\ast}_G \simeq 10^{-24}
\textrm{g}/\textrm{cm}^{3}$, we obtain
\begin{equation}
-5 \times 10^{-11} < f_R < 0 \quad \textrm{when} \quad
R/H_0^2 \sim 3 \times 10^5 .
\end{equation}

The case of the Earth experiments is more complicated. In this
case the spherically symmetric system has three regions: the
Earth, the atmosphere, and the beyond, with the mass densities
as follows.
\begin{equation}
\rho^{\ast}(r) = \left\{
\begin{array}{ll}
\rho^{\ast}_{\oplus} \simeq 5.5 \textrm{ g}/\textrm{cm}^3
& \textrm{ for } 0 < r < r_{\oplus} \, , \\
\rho^{\ast}_\textrm{atm} \simeq 10^{-3} \textrm{ g}/\textrm{cm}^3
& \textrm{ for } r_{\oplus} < r < r_\textrm{atm} \, , \\
\rho^{\ast}_G \simeq 10^{-24} \textrm{ g}/\textrm{cm}^3
& \textrm{ for } r > r_\textrm{atm} \, ,
\end{array}
\right.
\end{equation}
where the Earth radius $r_{\oplus} \simeq 6.4 \times 10^3$ km
and the thickness of the atmosphere $\Delta d_\textrm{atm} =
r_\textrm{atm} - r_{\oplus} \simeq $ 10--100 km.
With the following thin-shell condition satisfied,
\begin{equation} \label{eq:atm-thin-shell-parameter}
\frac{\Delta r_\textrm{atm}}{r_\textrm{atm}} \equiv
\frac{\kappa (\phi_G - \phi_\textrm{atm})}{6\beta \Phi_\textrm{atm}}
\ll 1 \, , \quad
\Phi_\textrm{atm} \equiv 
\frac{1}{6} \kappa^2 \rho^{\ast}_\textrm{atm} r^2_\textrm{atm} \, ,
\end{equation}
the scalar field profile is \cite{Khoury:2003rn}:
\begin{equation}
\phi (r) \simeq \left\{
\begin{array}{ll}
\phi_{\oplus} & \textrm{ for } 0 < r < r_{\oplus} \, , \\
\phi_\textrm{atm} & \textrm{ for } r_{\oplus} < r < r_\textrm{atm} \, , \\
\displaystyle - \left( \frac{\beta\kappa}{4\pi} \right)
\left( \frac{3\Delta r_{\oplus}}{r_{\oplus}} \right)
\frac{M_{\oplus} e^{-m_{G}(r-r_\textrm{atm})}}{r} + \phi_{G}
& \textrm{ for } r > r_\textrm{atm} \, ,
\end{array}
\right.
\end{equation}
\begin{equation}
\frac{\Delta r_{\oplus}}{r_{\oplus}} \equiv
\frac{\kappa (\phi_G - \phi_\textrm{atm})}{6\beta \Phi_{\oplus}}
\ll 1 \, , \quad
\Phi_{\oplus} \equiv 
\frac{1}{6} \kappa^2 \rho^{\ast}_{\oplus} r^2_{\oplus} \, ,
\end{equation}
where $\phi_{\oplus}$, $\phi_\textrm{atm}$ and $\phi_G$
respectively denote the locations of the effective potential
$V_\textrm{eff}$ minima in the three regions, $M_{\oplus}$ the
mass of the Earth, and $\Phi_{\oplus}$ and $\Phi_\textrm{atm}$
the Newtonian potentials:
$M_{\oplus} \simeq 6 \times 10^{24}\,$kg, %
$\Phi_{\oplus} 
\simeq 7 \times 10^{-10}$ and %
$\Phi_\textrm{atm} 
\simeq 10^{-13}$. %
Note that the shin-shell condition in Eq.\ (\ref{eq:atm-thin-shell-parameter}) 
automatically leads to $\Delta r_{\oplus}/r_{\oplus} \ll 1$.
The experimental bounds of the thin-shell parameters, 
$0 < \Delta r_i / r_i < (\Delta r_i / r_i)_\textrm{max} \ll 1$ 
for $i = \oplus , \textrm{atm}$,
then give upper bounds to $|f_R|$:
\begin{equation}
0 < - f_R (\kappa^2 \rho^{\ast}_G)
< 2 \Phi_i \cdot \left( \Delta r_i / r_i \right)_\textrm{max}
\ll \Phi_i \, , \quad i = \oplus , \textrm{atm} \, ,
\end{equation}
in the case where $| f_R (\kappa^2 \rho^{\ast}_\textrm{atm}) | \ll
| f_R (\kappa^2 \rho^{\ast}_G) |$ for $\rho^{\ast}_\textrm{atm} \gg
\rho^{\ast}_G$.

An essential experimental bound comes from the basic
requirement that the atmosphere has a thin shell, i.e., the
thickness of the thin shell should be smaller than that of the
atmosphere: $0 < \Delta r_\textrm{atm} / r_\textrm{atm} <
\Delta d_\textrm{atm} / r_\textrm{atm}$.
Taking the thickness of the atmosphere $\Delta d_\textrm{atm}
\simeq 50$ km, i.e.\ $\Delta d_\textrm{atm} / r_\textrm{atm}
\simeq 8 \times 10^{-3}$, we obtain a very stringent bound to
$f_R$:
\begin{equation}
-10^{-15} \lesssim f_R < 0 \quad \textrm{when} \quad
R/H_0^2 \sim 3 \times 10^5 .
\label{eq:fR constraint from atm}
\end{equation}


To sum up, from the thin-shell condition in the solar system we
have obtained a constraint on $R f_{RR}$ in Eq.\ (\ref{eq:fRR
constraint}) and a stringent upper bound to $|f_R|$ in Eq.\
(\ref{eq:fR constraint from atm}) for non-constant
$f(R)$.\footnote{Note that $f_R$ and $f_{RR}$ can vanish in the
case of constant $f(R)$ that is equivalent to the $\Lambda$CDM
model 
and therefore not taken into consideration when we investigate
$f(R)$ gravity for possible deviations from GR.}
They respectively come from the requirement of the existence of
the effective potential $V_\textrm{eff}(\phi,\rho^{\ast})$
minimum in the solar system and from the thin-shell condition
in the atmosphere, with the precondition that 
$|f| \ll R$ and $|f_R| \ll 1$ in the solar system, and 
$| f_R (\kappa^2 \rho^{\ast}_\textrm{atm}) | \ll | f_R (\kappa^2 \rho^{\ast}_G) |$.

The constraints on $f(R)$ we obtained can be conveniently applied to the $f(R)$ models where the functional forms of $f(R)$ are given. For demonstration, here we apply the constraint on $f_R$ in Eq.\ (\ref{eq:fR constraint from atm}) to the following widely considered models \cite{Tsujikawa:2009ku}: 
\begin{equation}
f(R) = - \lambda R_c f_1 (x), \quad x \equiv R/R_c, \quad \lambda,R_c>0 \, ;
\end{equation}
\begin{description}
\item (1) $f_1(x) = x^p$, $0<p<1 \,$;
\item (2) Hu and Sawicki \cite{Hu:2007nk}:
	$f_1(x) = x^{2n}/(x^{2n}+1)$, $n>0 \,$;
\item (3) Starobinsky \cite{Starobinsky:2007hu}:
	$f_1(x) = 1 - (1+x^2)^{-n}$, $n>0 \,$;
\item (4) Tsujikawa \cite{Tsujikawa:2007xu}:
	$f_1(x) = \tanh(x) \,$;
\item (5) Linder \cite{Linder:2009jz}:
	$f_1(x) = 1 - e^{-x} \,$.
\end{description}
Considering $\lambda \sim \mathcal{O}(1)$, $R_c \sim \kappa^2 \rho_c$ (where $\rho_c$ is the critical density at the present time) and $R/R_c \sim 10^5$, we find that the constraint in Eq.\ (\ref{eq:fR constraint from atm}) requires $p<10^{-10}$ in Model (1) and $n>1$ in Models (2) and (3), and it is well satisfied in Models (4) and (5). In addition, the constraint on $Rf_{RR}$ in Eq.\ (\ref{eq:fRR constraint}) is also satisfied in these models. Note that in these models and under the above consideration, the conditions $|f| \ll R$, $|f_R| \ll 1$ and 
$| f_R (\kappa^2 \rho^{\ast}_\textrm{atm}) | \ll | f_R (\kappa^2 \rho^{\ast}_G) |$ 
are satisfied, so that we can legitimately use the simple and stringent constraint on 
$f_R(\kappa^2 \rho^{\ast}_\textrm{G})$, i.e.\
the constraint in Eq.\ (\ref{eq:fR constraint from atm}).

\section{Conclusion} \label{sec:conclusion}

The cosmological tests and the local tests give essential
constraints on the deviation from GR in $f(R)$ gravity 
at different values of the Ricci scalar $R$, 
which correspond to the constraints 
at the cosmological scales at different epochs 
of the cosmic expansion history. Roughly speaking, CMB and BBN
stringently constrain the large-$R$ (early-time) behavior of
$f(R)$, the solar-system tests stringently constrain the
moderate-$R$ (middle-age) behavior, and the cosmological
observations about the late-time universe constrain the
small-$R$ (late-time) behavior.

With regard of the cosmic history, the constraints on 
the $f(R)$ modified gravity at the cosmological scales 
from the early times to the present are summarized as follows.
\begin{description}
  \item[$\bullet$ $z \gtrsim 10^3$:] The deviation from GR,
      such as $f_R$, should be small at early times when
      the redshift $z \gtrsim 10^3$, as required by the CMB
      and the BBN observations.
  \item[$\bullet$ $z \gtrsim 70$:] $0 < R f_{RR} < 2/5$
      when $R/H_0^2 \gtrsim 3 \times 10^5$ that roughly corresponds
      to the epoch $z \gtrsim 70$. This is a basic
      requirement in the chameleon mechanism for the solar
      system.
  \item[$\bullet$ $z \sim 70$:] $-10^{-15} < f_{R} < 0$
      when $R/H_0^2 \sim 3 \times 10^5$ that roughly corresponds to
      the time when $z \sim 70$. We obtain this stringent
      constraint from the thin-shell condition required in
      the solar-system test, particularly the test in the
      atmosphere.
  \item[$\bullet$ $z \sim \mathcal{O}(1)$:] The deviation
      from GR needs to be significant in order to explain
      the cosmic acceleration at the present epoch.
\end{description}

According to the above constraints, in the viable $f(R)$ models
of the late-time cosmic acceleration the deviation from GR should be small
when $z \gtrsim 70$ but become significant at the recent epoch.
The above constraints give simple, clear requirements one can
conveniently utilize to examine the viability of the $f(R)$ models 
with various functional forms of $f(R)$.

\begin{acknowledgments}
We thank the Dark Energy Working Group of the Leung Center for
Cosmology and Particle Astrophysics (LeCosPA). Gu is supported
by the Taiwan National Science Council (NSC) under Project No.\
NSC 98-2112-M-002-007-MY3 and Lin by Taiwan NSC under Project
No.\ NSC 98-2811-M-002-501.
\end{acknowledgments}

\newpage



\begin{thebibliography}{00}


\bibitem{Perlmutter:1998np}
  S.~Perlmutter {\it et al.}  [Supernova Cosmology Project Collaboration],
  Astrophys.\ J.\  {\bf 517}, 565 (1999)
  [arXiv:astro-ph/9812133].
\bibitem{Riess:1998cb}
  A.~G.~Riess {\it et al.}  [Supernova Search Team Collaboration],
  Astron.\ J.\  {\bf 116}, 1009 (1998)
  [arXiv:astro-ph/9805201].


\bibitem{Capozziello:2003tk}
  S.~Capozziello, S.~Carloni and A.~Troisi,
  Recent Res.\ Dev.\ Astron.\ Astrophys.\  {\bf 1}, 625 (2003)
  [arXiv:astro-ph/0303041].
\bibitem{Carroll:2003wy}
  S.~M.~Carroll, V.~Duvvuri, M.~Trodden and M.~S.~Turner,
  Phys.\ Rev.\  D {\bf 70}, 043528 (2004)
  [arXiv:astro-ph/0306438].
\bibitem{Nojiri:2003ft}
  S.~Nojiri and S.~D.~Odintsov,
  Phys.\ Rev.\  D {\bf 68}, 123512 (2003)
  [arXiv:hep-th/0307288].


\bibitem{Appleby:2007vb}
  S.~A.~Appleby and R.~A.~Battye,
  Phys.\ Lett.\  B {\bf 654}, 7 (2007)
  [arXiv:0705.3199 [astro-ph]].
\bibitem{Hu:2007nk}
  W.~Hu and I.~Sawicki,
  Phys.\ Rev.\  D {\bf 76}, 064004 (2007)
  [arXiv:0705.1158 [astro-ph]].
\bibitem{Starobinsky:2007hu}
  A.~A.~Starobinsky,
  JETP Lett.\  {\bf 86}, 157 (2007)
  [arXiv:0706.2041 [astro-ph]].
\bibitem{Nojiri:2007as}
  S.~Nojiri and S.~D.~Odintsov,
  Phys.\ Lett.\  B {\bf 657}, 238 (2007)
  [arXiv:0707.1941 [hep-th]].
\bibitem{Tsujikawa:2007xu}
  S.~Tsujikawa,
  Phys.\ Rev.\  D {\bf 77}, 023507 (2008)
  [arXiv:0709.1391 [astro-ph]].
\bibitem{Nojiri:2007cq}
  S.~Nojiri and S.~D.~Odintsov,
  Phys.\ Rev.\  D {\bf 77}, 026007 (2008)
  [arXiv:0710.1738 [hep-th]].
\bibitem{Cognola:2007zu}
  G.~Cognola, E.~Elizalde, S.~Nojiri, S.~D.~Odintsov, L.~Sebastiani and S.~Zerbini,
  Phys.\ Rev.\  D {\bf 77}, 046009 (2008)
  [arXiv:0712.4017 [hep-th]].
\bibitem{Linder:2009jz}
  E.~V.~Linder,
  Phys.\ Rev.\  D {\bf 80}, 123528 (2009)
  [arXiv:0905.2962 [astro-ph.CO]].
  

\bibitem{DeFelice:2010aj}
  A.~De Felice and S.~Tsujikawa,
  Living Rev.\ Rel.\  {\bf 13}, 3 (2010)
  [arXiv:1002.4928 [gr-qc]].


\bibitem{Amendola:2007nt}
  L.~Amendola and S.~Tsujikawa,
  Phys.\ Lett.\  B {\bf 660}, 125 (2008)
  [arXiv:0705.0396 [astro-ph]].
\bibitem{Tsujikawa:2008uc}
  S.~Tsujikawa, K.~Uddin, S.~Mizuno, R.~Tavakol and J.~Yokoyama,
  Phys.\ Rev.\  D {\bf 77}, 103009 (2008)
  [arXiv:0803.1106 [astro-ph]].
\bibitem{Serra:2009kp}
  P.~Serra, A.~Cooray, S.~F.~Daniel, R.~Caldwell and A.~Melchiorri,
  Phys.\ Rev.\  D {\bf 79}, 101301 (2009)
  [arXiv:0901.0917 [astro-ph.CO]].
\bibitem{Daniel:2009kr}
  S.~F.~Daniel, R.~R.~Caldwell, A.~Cooray, P.~Serra, A.~Melchiorri,
  Phys.\ Rev.\  D {\bf 80}, 023532 (2009)
  [arXiv:0901.0919 [astro-ph.CO]].
\bibitem{Guzik:2009cm}
  J.~Guzik, B.~Jain and M.~Takada,
  Phys.\ Rev.\  D {\bf 81}, 023503 (2010)
  [arXiv:0906.2221 [astro-ph.CO]].
\bibitem{Tsujikawa:2009ku}
  S.~Tsujikawa, R.~Gannouji, B.~Moraes and D.~Polarski,
  Phys.\ Rev.\  D {\bf 80}, 084044 (2009)
  [arXiv:0908.2669 [astro-ph.CO]].
\bibitem{Giannantonio:2009gi}
  T.~Giannantonio, M.~Martinelli, A.~Silvestri and A.~Melchiorri,
  JCAP {\bf 1004}, 030 (2010)
  [arXiv:astro-ph/0909.2045].
\bibitem{Daniel:2010ky}
  S.~F.~Daniel, E.~V.~Linder, T.~L.~Smith, R.~R.~Caldwell, A.~Cooray, A.~Leauthaud and L.~Lombriser,
  Phys.\ Rev.\  D {\bf 81}, 123508 (2010)
  [arXiv:1002.1962 [astro-ph.CO]].
\bibitem{Bean:2010zq}
  R.~Bean and M.~Tangmatitham,
  Phys.\ Rev.\  D {\bf 81}, 083534 (2010)
  [arXiv:1002.4197 [astro-ph.CO]].
\bibitem{LinGuChen:2010}
  W.-T.~Lin, Je-An~Gu and P.~Chen,
  arXiv:1009.3488 [astro-ph.CO].


\bibitem{Chiba:2006jp}
  T.~Chiba, T.~L.~Smith and A.~L.~Erickcek,
  Phys.\ Rev.\  D {\bf 75}, 124014 (2007)
  [arXiv:astro-ph/0611867].
\bibitem{Faulkner:2006ub}
  T.~Faulkner, M.~Tegmark, E.~F.~Bunn and Y.~Mao,
  Phys.\ Rev.\  D {\bf 76}, 063505 (2007)
  [arXiv:astro-ph/0612569].
\bibitem{Capozziello:2007eu}
  S.~Capozziello and S.~Tsujikawa,
  Phys.\ Rev.\  D {\bf 77}, 107501 (2008)
  [arXiv:gr-qc/0712.2268].
\bibitem{Bisabr:2009ee}
  Y.~Bisabr,
  Phys.\ Lett.\  B {\bf 683}, 96 (2010)
  [arXiv:0907.3838 [gr-qc]].


\bibitem{Khoury:2003aq}
  J.~Khoury and A.~Weltman,
  Phys.\ Rev.\ Lett.\  {\bf 93}, 171104 (2004)
  [arXiv:astro-ph/0309300].
\bibitem{Khoury:2003rn}
  J.~Khoury and A.~Weltman,
  Phys.\ Rev.\  D {\bf 69}, 044026 (2004)
  [arXiv:astro-ph/0309411].


\bibitem{Will:2005va}
  C.~M.~Will,
  Living Rev.\ Rel.\  {\bf 9}, 3 (2005)
  [arXiv:gr-qc/0510072].

\end{thebibliography}
\end{document}